\documentclass[11pt]{article}

\usepackage[margin=1in]{geometry}
\usepackage{lmodern}
\usepackage{microtype}
\usepackage{setspace}
\usepackage{amsmath,amssymb,mathtools,bm}
\usepackage{amsthm}
\numberwithin{equation}{section}
\usepackage{booktabs}
\usepackage{tabularx}
\usepackage{array}
\usepackage{enumitem}
\usepackage[round,authoryear]{natbib}
\usepackage{hyperref}
\hypersetup{
  colorlinks=true,
  linkcolor=black,
  citecolor=black,
  urlcolor=black,
  pdftitle={Mean-Tilted Intervals: Regression and Dynamic Models},
  pdfauthor={Antonio De Leon, Raquel Prado, and Bruno Sanso}
}

\newcommand{\E}{\mathbb{E}}
\newcommand{\Normal}{\mathcal{N}}
\newcommand{\Exp}{\mathrm{Exp}}
\newcommand{\GIG}{\mathrm{GIG}}
\newcommand{\ind}{\mathbf{1}}
\newcommand{\diag}{\mathrm{diag}}

\newcommand{\mat}[1]{\bm{#1}}

\title{\vspace{-0.6cm}Mean-Tilted Intervals:\\
Regression and Dynamic Models}
\author{Antonio De Leon$^{1}$ \and Raquel Prado$^{1}$ \and Bruno Sans\'o$^{1}$}
\date{\small $^{1}$ Department of Statistics, University of California, Santa Cruz\\
\vspace{0.25em}August 2026}

\begin{document}
\maketitle
\vspace{-1.25em}

\begin{abstract}
\noindent
Mean-tilted intervals (MTI) provide a fixed-content family of interval targets
indexed by retained-mean balance. This paper develops generalized-Bayes
computation for using those targets in regression and time-varying endpoint
models after content and tilt have been specified. The update combines a prior
with an exponentiated residual-product check loss, yielding uncertainty
summaries for endpoint functions rather than a response-distribution model.

A pseudo-asymmetric-Laplace normal-exponential representation gives
generalized-inverse-Gaussian latent-scale updates and Gaussian conditional
updates for one root when the other is fixed. The same representation gives
deterministic ECM equations for generalized-posterior modes. We present ridge
endpoint regression, a conditionally Gaussian shrinkage construction for the
zero-tilt case, deterministic fixed-feature readouts, and dynamic linear
endpoint states updated by alternating forward-filtering backward-sampling
steps. These methods extend the MTI target from iid intervals to regression and
dynamic settings; tolerance certification and response prediction require
separate calibration or response models.
\end{abstract}

\section{Introduction}
\label{sec:introduction}

Intervals with the same probability content can represent different statistical
targets. Equal-tailed, mean-preserving, and shortest-contiguous intervals share
content but differ in endpoint placement. The companion tolerance paper
develops this geometry and uses it to define scan-calibrated tolerance actions.
The present paper focuses on a different question: once a fixed-content target
has been specified, how can endpoint functions be estimated and summarized in
regression and time-varying settings?

The starting point is the residual-product criterion introduced by
\citet{PouplinEtAl2024RQR}. For ordered endpoints \(a<b\), the product
residual \((y-a)(y-b)\) is negative exactly when \(y\) lies inside the interval.
With a check loss applied to this scalar residual, the unrestricted population
score equations identify the mean-preserving interval (MPI). MTI generalizes
MPI by adding a fixed retained-mean tilt \(\delta\), so that different
admissible fixed-content windows are represented by different tilt values.

This paper treats MPI and MTI as loss-defined targets. For a fixed content
\(c\), tilt \(\delta\), and learning rate \(\omega_{\mathrm R}\), a
generalized posterior is obtained by updating a prior with
\(\exp\{-\omega_{\mathrm R}L_{c,\delta,n}\}\)
\citep{BissiriHolmesWalker2016GeneralBayes}. This update is useful for
regularization and uncertainty summaries for endpoint functions. Because the
update targets endpoints, posterior endpoint draws should be interpreted as
uncertainty summaries for the interval functional. Repeated-sampling tolerance
statements require a separate calibration layer.

The contributions are threefold. First, we formulate static endpoint regression
for MPI and fixed-tilt MTI, including the root-label symmetry and the
midpoint-gap summaries needed for interpretation. Second, we derive a
pseudo-asymmetric-Laplace augmentation that gives Gaussian conditional root
updates, generalized-inverse-Gaussian latent-scale updates, and a deterministic
ECM mode algorithm. Third, we show how the same fixed-target computation extends
to deterministic feature readouts and to dynamic linear endpoint states through
alternating root-specific smoothing updates.

The formal setting is deliberately bounded. Nonzero tilt is treated as fixed before
fitting or selected by an external rule. The conditionally Gaussian shrinkage
construction is stated for the MPI case unless an additional propriety argument
is supplied. Dynamic endpoint models describe time-varying interval-root
functionals under the chosen loss and prior; they do not inherit the iid
tolerance certificate studied in the companion paper.

\section{Fixed-Content Targets and Generalized Bayes}
\label{sec:targets}

Let \(Y\) have a continuous distribution with finite mean \(\mu\), and let
\(c\in(0,1)\) denote the desired interval content. For endpoints \(a<b\), define
\[
e(a,b;y)=(y-a)(y-b),
\qquad
\rho_c(r)=r\{c-\ind(r<0)\}.
\]
The MPI loss is
\[
\ell_c^{\mathrm{MPI}}(a,b;y)=\rho_c\{e(a,b;y)\}.
\]
At an unrestricted regular population stationary point, the score equations
give
\[
\Pr(a<Y<b)=c,
\qquad
\E(Y\mid a<Y<b)=\mu.
\]
Thus MPI selects the content-\(c\) interval whose retained mean equals the full
mean.

MTI replaces zero retained-mean balance by a fixed offset. In midpoint and
half-width coordinates \(a=m-h\), \(b=m+h\), the fixed-tilt loss can be written
\[
\ell_{c,\delta}(m,h;y)
=\rho_c\{(y-m)^2-h^2\}-2c\delta(y-m).
\]
At a regular unrestricted stationary point,
\[
\Pr(|Y-m|<h)=c,
\qquad
\E(Y\mid |Y-m|<h)=\mu+\delta .
\]
The tilt \(\delta\) is part of the target specification. Treating it as an
ordinary unknown inside the same posterior would average over distinct interval
targets.

For data \(D_n=\{(x_i,y_i):i=1,\ldots,n\}\), write the two raw root functions as
\(\eta_1(x;\theta_1)\) and \(\eta_2(x;\theta_2)\). Given fixed
\((c,\delta,\omega_{\mathrm R})\), the generalized posterior is
\begin{equation}
\pi(\theta_1,\theta_2\mid D_n)
\propto
\exp\{-\omega_{\mathrm R}L_{c,\delta,n}(\theta_1,\theta_2)\}
\pi_0(\theta_1,\theta_2).
\label{eq:generalized-posterior}
\end{equation}
The learning rate controls the scale of the loss update. It is distinct from
the content \(c\), the tilt \(\delta\), and any response-distribution parameter.

\section{Static Endpoint Regression}
\label{sec:static-regression}

In static linear endpoint regression, the raw roots are
\[
\eta_{1i}=x_i^\top\beta_1,\qquad
\eta_{2i}=x_i^\top\beta_2.
\]
The loss is symmetric in the two roots. Lower and upper endpoints are therefore
reported after sorting the fitted root values,
\[
L_i=\min(\eta_{1i},\eta_{2i}),\qquad
U_i=\max(\eta_{1i},\eta_{2i}),\qquad
W_i=U_i-L_i.
\]
This sorting is applied to endpoint functionals, not to individual regression
coefficients.

The midpoint-gap coordinates clarify which summaries are invariant to root
labels:
\[
\beta_m=\frac{\beta_1+\beta_2}{2},
\qquad
\beta_s=\frac{\beta_2-\beta_1}{2}.
\]
At a design row \(x\),
\[
M(x)=x^\top\beta_m,\qquad
S(x)=x^\top\beta_s,
\]
and
\[
L(x)=M(x)-|S(x)|,\qquad
U(x)=M(x)+|S(x)|,\qquad
W(x)=2|S(x)|.
\]
The vector \(\beta_m\) is the coefficient vector for the geometric midpoint.
The signed vector \(\beta_s\) is an orientation coordinate for the gap. A direct
lower-endpoint and upper-endpoint coefficient interpretation requires a design
region where one root remains below the other.

Ridge priors give a stable default for fixed-dimensional endpoint regression.
Let
\[
\beta_j\sim\Normal(b_{0j},P_{0j}^{-1}),\qquad j=1,2,
\]
with proper precision matrices \(P_{0j}\). Proper Gaussian priors are also
useful for fixed nonzero tilt because the tilt term is linear in the root
parameters and the Gaussian prior has finite moment-generating functions in
finite dimensions.

\section{Pseudo-AL Computation}
\label{sec:pseudo-al}

The product-residual check loss admits a normal-exponential representation
analogous to the asymmetric-Laplace construction used in Bayesian quantile
regression \citep{yu2001bayesian,KozumiKobayashi2011Gibbs}. Let
\[
\sigma_{\mathrm R}=\omega_{\mathrm R}^{-1},\qquad
\xi_c=\frac{1-2c}{c(1-c)},\qquad
\phi_c=\frac{2}{c(1-c)}.
\]
For \(e_i=(y_i-\eta_{1i})(y_i-\eta_{2i})\),
\begin{align}
&\frac{c(1-c)}{\sigma_{\mathrm R}}
\exp\{-\rho_c(e_i)/\sigma_{\mathrm R}\}\notag\\
&\quad=
\int_0^\infty
\Normal(e_i\mid \xi_cv_i,\phi_c\sigma_{\mathrm R}v_i)
\Exp(v_i\mid\mathrm{rate}=1/\sigma_{\mathrm R})\,dv_i .
\label{eq:pseudo-al}
\end{align}
This identity augments the generalized-posterior factor in
\eqref{eq:generalized-posterior}. The resulting kernel is defined on product
residuals and should be interpreted as a loss factor for endpoints; a sampling
density for \(Y_i\) would require an additional response model.

The latent-scale full conditional is
\[
v_i\mid\cdot\sim
\GIG\left\{
\frac12,\
\frac{1}{2\sigma_{\mathrm R}c(1-c)},\
\frac{c(1-c)e_i^2}{2\sigma_{\mathrm R}}
\right\}.
\]
Conditional on \((v_i)\) and one root, the other root has a Gaussian update.
For example, holding \(\eta_2\) fixed, define
\[
\mat A_1=\diag(y_i-\eta_{2i})\mat X,\qquad
z_1=y\circ(y-\eta_2)-\xi_c v.
\]
With weight matrix
\[
\mat W_v=\diag\{(\phi_c\sigma_{\mathrm R}v_i)^{-1}\},
\]
the canonical parameters for \(\beta_1\) are
\[
\mat Q_1=P_{01}+\mat A_1^\top\mat W_v\mat A_1,
\qquad
h_1=P_{01}b_{01}+\mat A_1^\top\mat W_v z_1.
\]
The conditional update is
\[
\beta_1\mid\cdot\sim \Normal(\mat Q_1^{-1}h_1,\mat Q_1^{-1}).
\]
The update for \(\beta_2\) is obtained by exchanging root labels.

For fixed nonzero tilt, the conditional precision is unchanged and the
canonical vector receives the additive shift
\[
h_1^{\mathrm{MTI}}
=h_1+\omega_{\mathrm R}c\mat X^\top\delta ,
\]
with the same form for the second root. This is the main computational
convenience of the fixed-tilt construction.

The same augmentation yields an ECM mode calculation
\citep{DempsterLairdRubin1977EM,MengRubin1993ECM}. For \(e_i\ne0\),
\[
E(v_i^{-1}\mid\cdot)=\frac{1}{c(1-c)|e_i|}.
\]
Using these inverse-scale moments, the conditional maximization for each root
is a Gaussian ridge solve. The numerical implementation uses objective
backtracking and deterministic multi-start selection, because the joint
criterion is not globally quadratic in the two roots.

\section{Fixed Features and Time-Varying Endpoints}
\label{sec:extensions}

Deterministic feature readouts use the static regression equations with a
larger fixed design matrix. For example, a frozen echo-state or deep
echo-state representation supplies deterministic nonlinear features
\(h_i\), after which the root functions are
\[
\eta_{1i}=h_i^\top\beta_1,\qquad
\eta_{2i}=h_i^\top\beta_2.
\]
The recurrent feature construction is a representation of the predictors. The
generalized posterior remains the fixed-target endpoint update in
\eqref{eq:generalized-posterior}. Hyperparameter selection for the feature map
must be recorded as part of the validation design.

Time-varying endpoint models replace static coefficients by root-specific state
trajectories. A basic dynamic linear specification is
\[
\eta_{jt}=f_t^\top\theta_{jt},\qquad
\theta_{jt}=G_t\theta_{j,t-1}+\varepsilon_{jt},\qquad j=1,2,
\]
with Gaussian state innovations. The stacked prior over both root paths is
Gaussian. The product residual makes the joint augmented log density quartic in
the two paths together, but conditional on one complete root path the other path
has a linear Gaussian observation equation. Alternating root-specific
forward-filtering backward-sampling steps therefore gives a natural blocked
sampler for the specified fixed-target generalized posterior
\citep{GoncalvesMigonBastos2020DQLM}.

Discount factors and component scales affect the state prior and hence the
posterior target. Fixed discount templates can be treated as part of the model
specification. Adaptive discount or scale selection changes the inferential
target unless the selection rule is stated as part of a larger model or as a
separate sensitivity analysis.

\section{Uncertainty Calibration and Diagnostics}
\label{sec:diagnostics}

The generalized posterior in \eqref{eq:generalized-posterior} is centered at a
loss-defined target under standard local conditions, but its curvature
covariance need not equal the frequentist sandwich covariance. In fixed
dimension, open-faced-sandwich adjustment \citep{Shaby2014OpenFacedSandwich}
can be used as a post-processing calibration for differentiable coefficient or
endpoint summaries when consistent curvature and score-variance estimates are
available. This adjustment calibrates generalized-posterior uncertainty; it is
separate from any response-distribution model used for prediction.

For MCMC calculations, reporting should include the number of chains, warmup and
retained draws, convergence diagnostics, effective sample sizes, and any
root-label handling. For ECM calculations, reporting should include the
selected content and tilt, the final objective value, the number of iterations,
the stopping rule, and stability checks across starting values. For deterministic
feature and time-varying endpoint models, the feature construction, state prior,
and validation protocol should be stated before empirical comparisons are made.

Empirical evaluation must match the target. Endpoint regression can be assessed
through held-out MTI loss, empirical content, interval width, and recovery of
known endpoint functions in simulation. Tolerance validation requires a
content-confidence action, such as the TCSP construction in the companion
tolerance paper. Forecast evaluation requires a response-distribution model or a
clearly defined predictive interval rule.

\section{Discussion}
\label{sec:discussion}

MTI provides a way to specify fixed-content interval placement through
retained-mean balance. The generalized-Bayes construction developed here turns a
chosen fixed-content target into regularized endpoint regression, deterministic
feature readouts, and time-varying endpoint models. The pseudo-AL augmentation
gives practical Gibbs and ECM computations while preserving the distinction
between a loss update and a response likelihood.

The main limitation is also the main organizing principle: content, endpoint
uncertainty, and tolerance confidence are different statistical statements.
Nonzero tilt should be fixed or selected by a separate rule before fitting.
Regression and time-varying endpoint models require their own validation before
claims about repeated-sampling tolerance behavior are made. This separation
keeps the computation extensible without overstating the guarantees attached to
the interval targets.

\clearpage
\bibliographystyle{plainnat}
\bibliography{refs}

\end{document}